\documentclass{article}  
\usepackage{bigsky2007}
\usepackage{graphicx}
\newcommand{\pt}{p$_{_{T}}$}
\newcommand{\snn}{$\sqrt{s_{NN}}=$}

\frompage{000} \topage{000}                                              
\def\rightmark{Identified particles at high-\pt}
\def\leftmark{R.S.Hollis {\it et al.}}
 
\title{Identified particle measurements at large transverse\\momenta in Cu+Cu collisions at RHIC.} 
\authors{
{Richard S Hollis$^1$ for the STAR Collaboration
}\\[2.812mm]
{\normalsize
\hspace*{-8pt}$^1$ University of Illinois at Chicago, \\ 
Chicago, IL, 60607, USA\\[0.2ex] 
}}
 
\abstract{Differential measurements of various particle species over
an extended momentum range provide a sensitive experimental tool for
investigation of energy loss mechanisms in the medium created in
nucleus-nucleus collisions at RHIC. In these proceedings, a systematic
study of transverse momentum spectra for charged pions, protons and
antiprotons from Cu+Cu data at \snn~200GeV as a function of collision
centrality will be presented.  Such systematic studies provide additional
insights into the interplay between fragmentation and non-fragmentation
contributions to the particle production. To investigate system size
effects on energy loss, a comparison of top energy results for Cu+Cu
and Au+Au collision systems are made.}

\keyword{Cu+Cu, hadronization, high-\pt} 
\PACS{25.75.-q} 
 
\begin{document}
 
\maketitle
\setcounter{page}{1}

\section{Introduction}\label{sec:intro}
Nucleus-nucleus collisions at the Relativistic Heavy-Ion Collider (RHIC)
have produced charged particle measurements that show two dramatic
effects which cannot be explained in terms of a simple extrapolation
from p+p collision data.  Such effects, for example high-\pt~hadron
suppression in central Au+Au
collisions~\cite{cite:STAR_SpectralSuppression,cite:STAR_BackToBack}, have
been attributed to partonic energy loss in the hot, dense medium created
in the ultra-relativistic heavy-ion collisions~\cite{cite:STAR_dAuBackToBack}.
In these proceedings, a systematic study of this and other effects will be
presented in terms of identified particle measurements in Cu+Cu collisions
at \snn~200~GeV.

In elementary collisions, hard partonic scatterings are known to produce
jets of particles originating from a fragmenting high-\pt~quark or gluon.
Spectral distributions of particles in transverse momentum from such
interactions are measured from experiment and are reasonably well
understood from NLO pQCD calculations~\cite{cite:STAR_Id200dAupp}.  These
hard scatterings are still present in the heavy-ion data, but resulting
distributions are found to be modified due to medium interactions.  Thus,
understanding modifications to the high-\pt~particle distributions can lead
to qualitative conclusions on the energy loss mechanisms within the medium.
As the unmodified (vacuum fragmentation) distributions of high-\pt~particles
are known from elementary p+p collisions, comparative analysis provides a
distinct advantage for hard sector studies.

Deviation from the p+p expectation could be expressed through the nuclear
modification factor, R$_{AA}$ (Eqn.~\ref{eqn:RAA}).  Eqn.~\ref{eqn:RAA}
reflects the expectation that the high-\pt~particle distributions should
follow directly from p+p collisions, scaled by the number of binary
(nucleon-nucleon) collisions, N$_{bin}$. For perfect binary scaling, a
value of R$_{AA}=1$ would be expected.  For R$_{AA}<1$ the spectra is
referred to as {\it suppressed}, whilst R$_{AA}>1$ corresponds to an
{\it enhancement}.  

\begin{equation}
   R_{AA} = \frac{\sigma_{NN}^{inel}}{N_{bin}^{AA}} \frac{d^{2}N_{AA}/dydp_{_{T}}}{d^{2}\sigma_{pp}/dydp_{_{T}}} \,. 
\label{eqn:RAA}
\end{equation}
\vspace{5pt}

\noindent Two experimentally observed modifications are noted {\it Cronin}
enhancement~\cite{cite:STAR_Id200dAupp,cite:CroninEffect} (present in
d+Au and peripheral
Au+Au collisions) and {\it jet quenching} (in central Au+Au collisions).
The former is believed to be due to multiple nucleon scattering within
the nucleus and the latter due to interactions with the hot, dense
medium.  Neither of these are clearly understood thus require further
experimental and theoretical study.

The study of high-\pt~particles is clearly an important step toward
understanding medium effects in heavy-ion collisions.  More information
can be gleaned, however, by connecting such studies with which type of 
partons are propagating through the medium.  Such information cannot be
obtained in the heavy-ion environment on an event-by-event basis or on
the jet level, but can
be done on a statistical basis by considering an ensemble of events.  To
tag, statistically, gluon or quark partons, one can use protons (baryons)
or pions (mesons).  Pion production is readily attributed to quark jets
by considering the breaking of strings.  It is energetically favorable to
produce a $q\bar{q}$ pair rather than a $qq\bar{q}\bar{q}$ needed for
baryon production in this regime.  Gluon jets are known to have a softer
fragmentation function and thus, for the same energy as a given quark jet,
need to produce more particles or heavier particles.  Identified proton and
$\pi$ measurements from p+p collisions concur with this
picture~\cite{cite:STAR_Id200dAupp,cite:STAR_Id200AuAu}.  Thus, identified
particle measurements can then be used to obtain information of gluon and
quark propagation through the medium and to probe the color-charge
differences of energy loss~\cite{cite:STAR_Id200AuAu,cite:Theory_Vitev,cite:STAR_62Id,cite:STAR_BedangaProc}.

In order to systematically test the effects observed in Au+Au collisions
at RHIC, several test experiments have been performed.  The first one was
the collision of d+Au nuclei to probe nuclear effects on the particle
production without the hot, dense medium of Au+Au collisions.  The second
test, the focus of these proceedings, is the collision of much smaller
nuclei, Cu+Cu, to gain a greater understanding of {\it peripheral} collisions. 
The size of Cu+Cu nucleus are ideally suited to explore the turn-on of
the high-\pt~suppression and bridge the gap between d+Au and peripheral
Au+Au data in terms of system size.

\section{Analysis Methods}\label{sec:analysis}

\begin{figure}[t]
\begin{minipage}[t]{.45\textwidth}
\centerline{\epsfxsize=2.4in\epsfbox{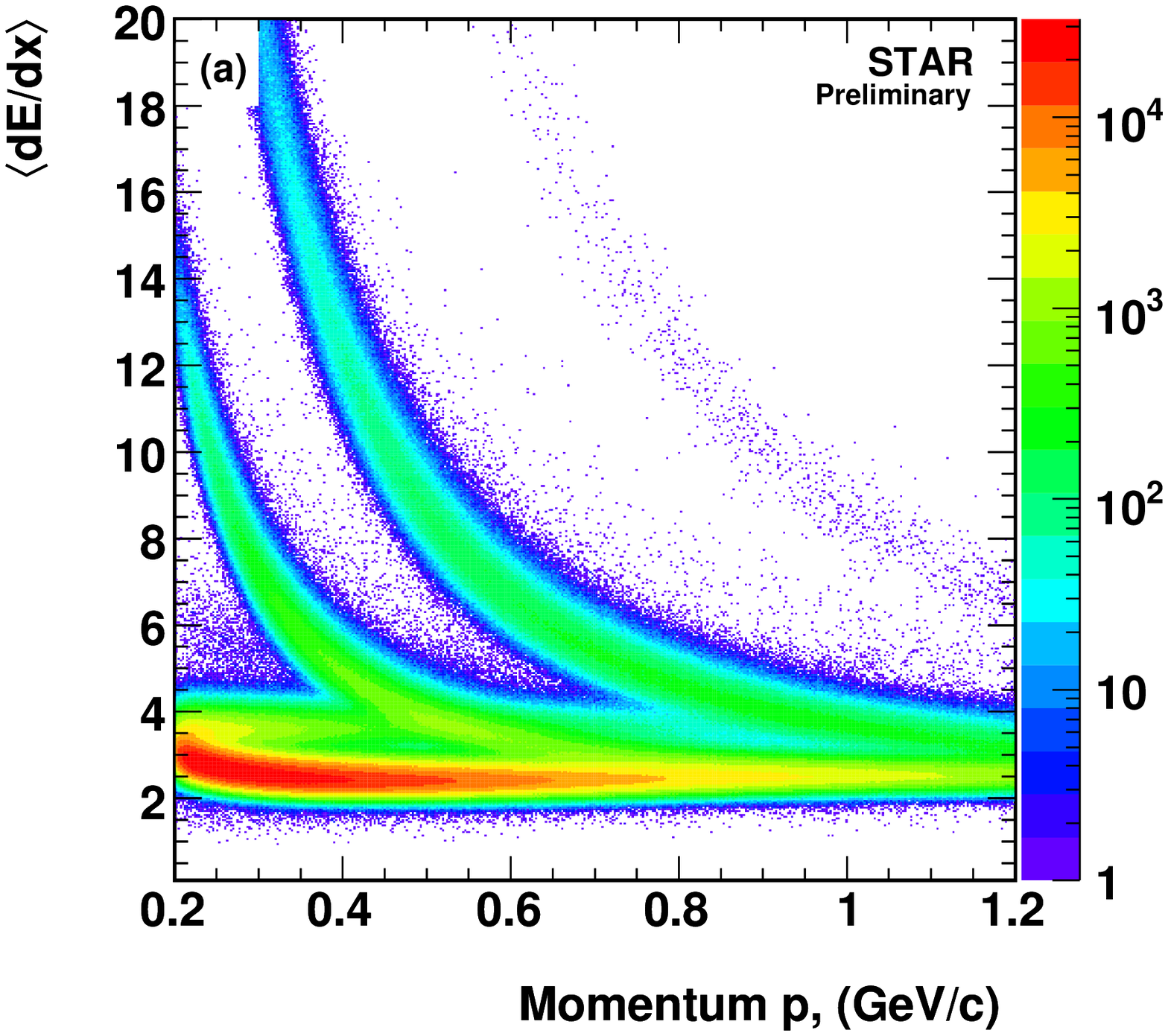}}
\end{minipage}
\hspace{20pt}
\begin{minipage}[t]{.45\textwidth}
\centerline{\epsfxsize=2.4in\epsfbox{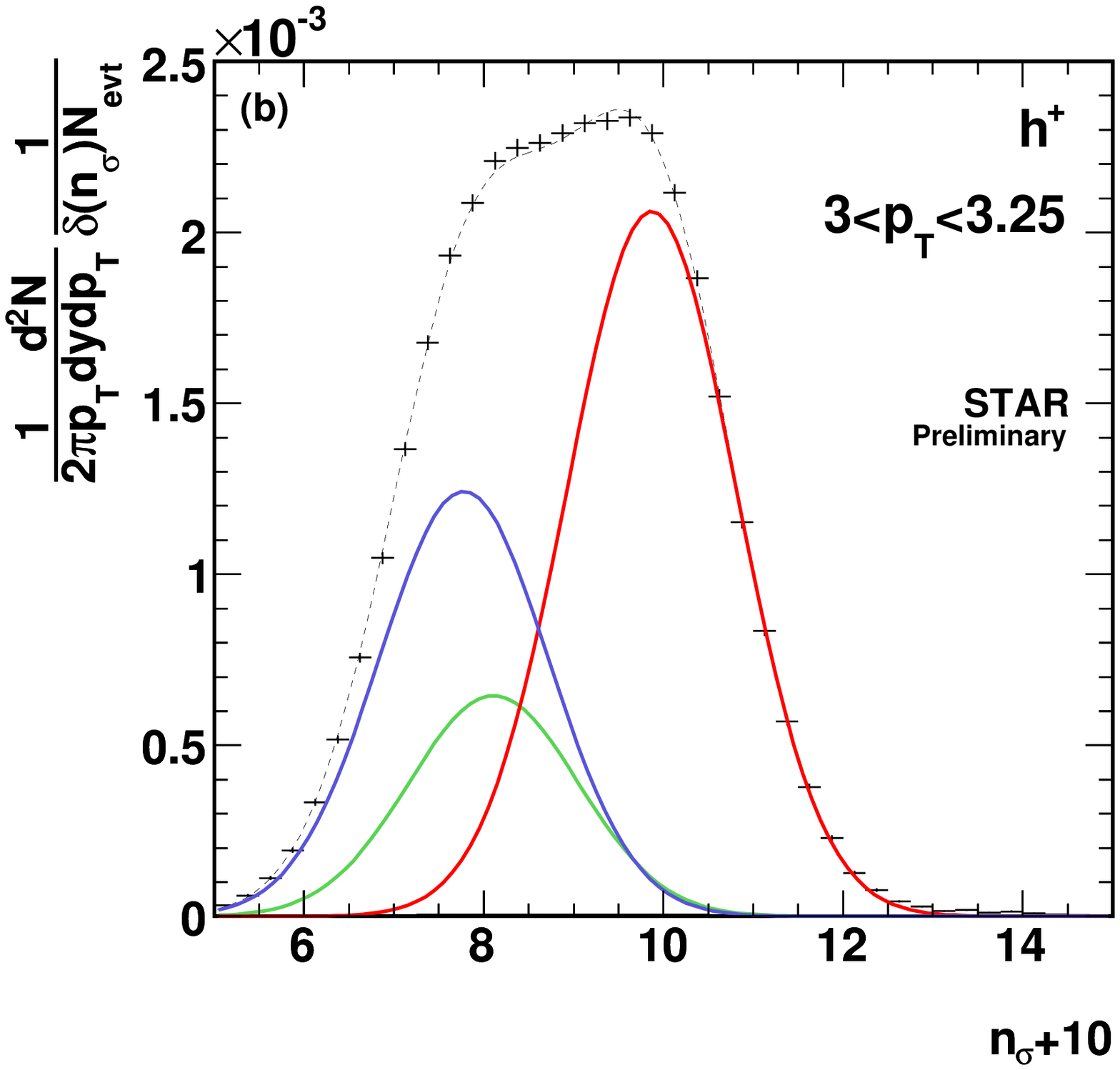}}
\end{minipage}
\caption{Panel (a) illustrates the particle species bands at low-\pt~from
Cu+Cu collisions at \snn~200GeV.  $\pi$, K and proton bands are clearly
separated.  Panel (b) shows the ionization energy loss for particles with
high transverse momenta ($3.00<$\pt$<3.25$~GeV$/c$), the non-gaussian shape
is used to extract the $\pi$ and proton yields.
\label{fig:PIDBands}}
\end{figure}

Particle identification at low-\pt~is attained by use of the ionization
energy loss in the TPC~\cite{cite:STAR_TPC_NIM}.  For low momentum particles
$0.2<$\pt$<1.2$~GeV$/c$, a clear mass separation is observed,
Fig.~\ref{fig:PIDBands}a, allowing the identification of $\pi^{\pm}$,
K$^{\pm}$ and (anti)protons.  In the intermediate-\pt~region
($1<$\pt$<3$~GeV$/c$) the TPC is no longer directly usable, as all particles,
independent of mass, are minimum ionizing.  In this region, the Time of Flight
(ToF) system is employed to identify particle species up to 3~GeV$/c$ in
transverse momentum.  Clear particle bands appear for ToF data in this 
kinematic region and allow for a direct identification of particle species.
At higher momenta, where charged pions, kaons and protons and anti-protons
are not separable into clear bands, the relativistic rise of the ionization
energy loss in the TPC is exploited to statistically identify particles.
Here, for a given slice in transverse momentum, a distinctly
non-single-gaussian shape can be seen, Fig.~\ref{fig:PIDBands}b,
representing the different energy losses of $\pi$, K and protons.
The quantity used to express the energy loss is a normalized distribution,
$n_{\sigma}$ defined in Eqn.~\ref{eqn:Nsigma}, which accounts for the 
theoretical expectation ($B_{\pi}$) and the resolution of the TPC for pions.

\begin{equation}
n_{\sigma}={\rm log}((dE/dx)/(B_{\pi}))/\sigma_{\pi}
\label{eqn:Nsigma}
\end{equation}

\noindent We fit this distribution with a triple gaussian (one per particle),
where the widths and single-gaussian centroids are constrained.  The yield
of charged kaons is obtained by two complementary methods: bin counting and a
direct measurement of K$^{0}_{S}$.  Once obtained, this kaon yield is fixed
in the fitting procedure of $n_{\sigma}$.  Further details of the analysis
techniques can be found in Refs.~\cite{cite:STAR_Id200dAupp,cite:STAR_ToF}.

\section{Results}\label{sec:results}

\begin{figure}[!b]
\begin{minipage}[t]{.45\textwidth}
\centerline{\epsfxsize=2.4in\epsfbox{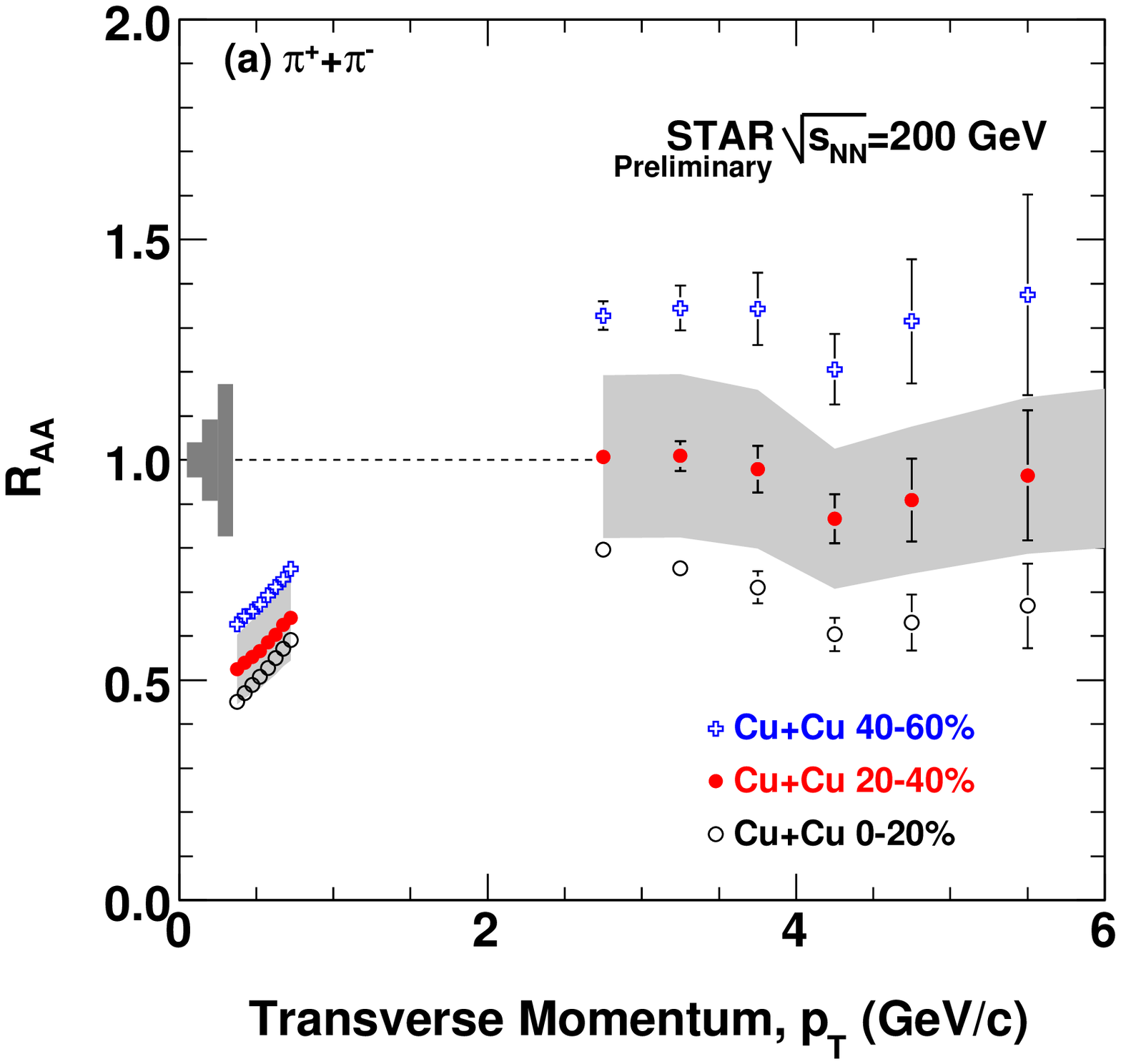}}
\end{minipage}
\hspace{20pt}
\begin{minipage}[t]{.45\textwidth}
\centerline{\epsfxsize=2.4in\epsfbox{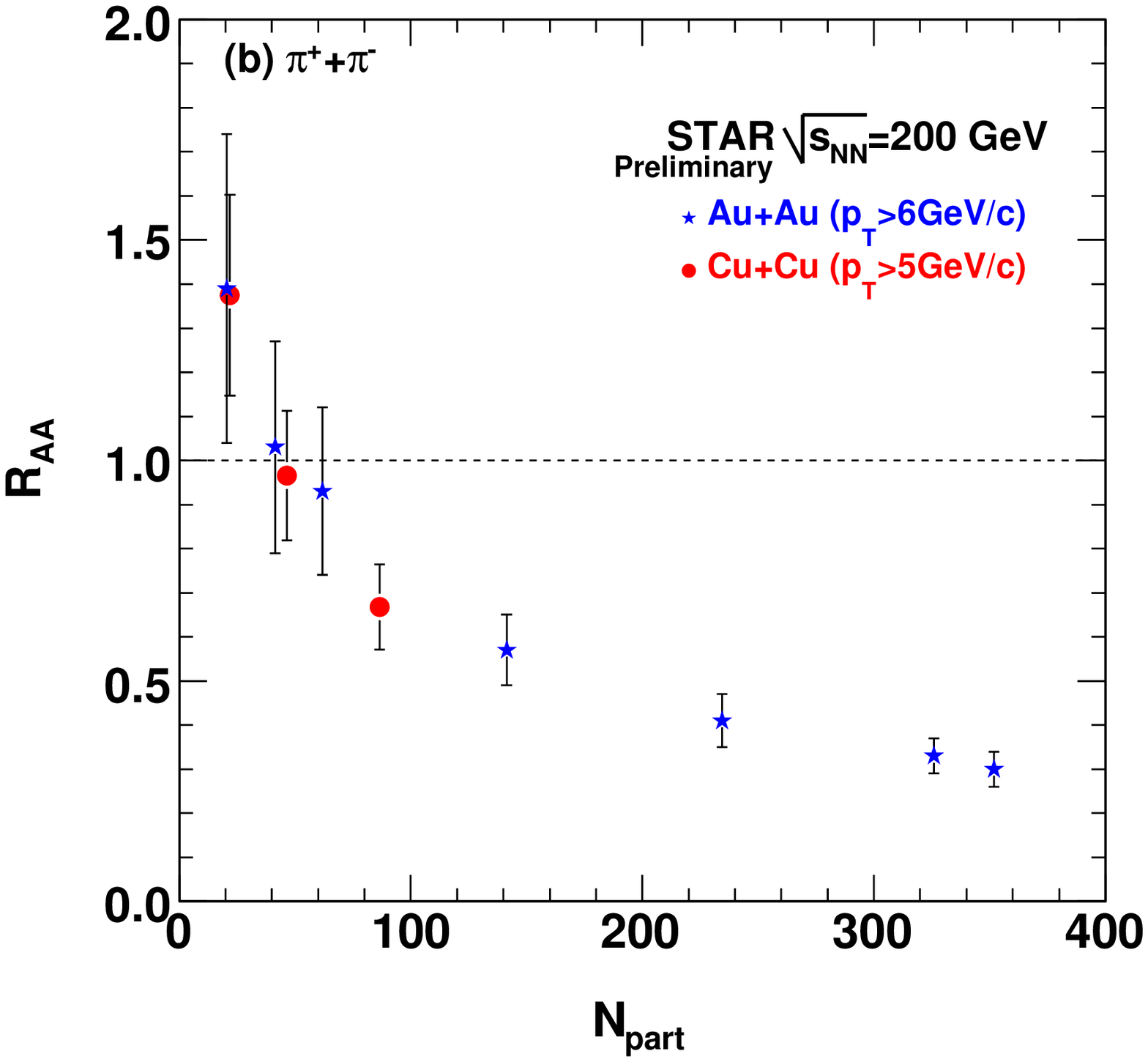}}
\end{minipage}
\caption{Panel (a) shows the transverse momentum dependence of
$\pi^{+}+\pi^{-}$ $R_{AA}$ in Cu+Cu collisions at \snn~200GeV for a
central, mid-central and a mid-peripheral centrality bin.  The shaded
band over the data points represent the systematic uncertainty in the
Cu+Cu data, the grey shaded band, at $R_{AA}=1$, illustrates the scale
uncertainty on the measurement from $N_{coll}$ for each Cu+Cu centrality
bin.  Panel (b) shows the centrality dependence of the energy loss for
the same data, averaged for \pt$>6$~GeV$/c$.
\label{fig:PionRAA}}
\end{figure}

As discussed in the introduction, the medium modification to the spectra
can be described by the nuclear modification factor, R$_{AA}$, defined
in Eqn.~\ref{eqn:RAA}.  The modification to the pion spectra for Cu+Cu
data, Fig.~\ref{fig:PionRAA}a, show a suppression relative to the
expectation from binary collision scaled p+p data at high-\pt~for the
most central data.  The most peripheral collisions show an enhancement
in the high-\pt~region.  These results are commensurate with the Au+Au
results at an equivalent number of participants, fitting consistently
into the $N_{part}$ systematics, Fig.~\ref{fig:PionRAA}b.  The smooth
dependence of the nuclear modification factor could be explained as a
consequence of medium induced energy loss of partons traversing the hot,
dense medium.  For the smaller systems sizes, either peripheral Au+Au or
Cu+Cu data, the path length traversed is smaller (on average) than for
the larger system (central Au+Au).  A smaller energy loss is thus
predicted~\cite{cite:Theory_Vitev} as observed in the data.  This
detailed prescription determined that R$_{AA}$ scales with the number
of participants, a phenomenon that is evident in other aspects of the 
data, discussed later in these proceedings.

\begin{figure}[ht]
\begin{minipage}[t]{.45\textwidth}
\centerline{\epsfxsize=2.4in\epsfbox{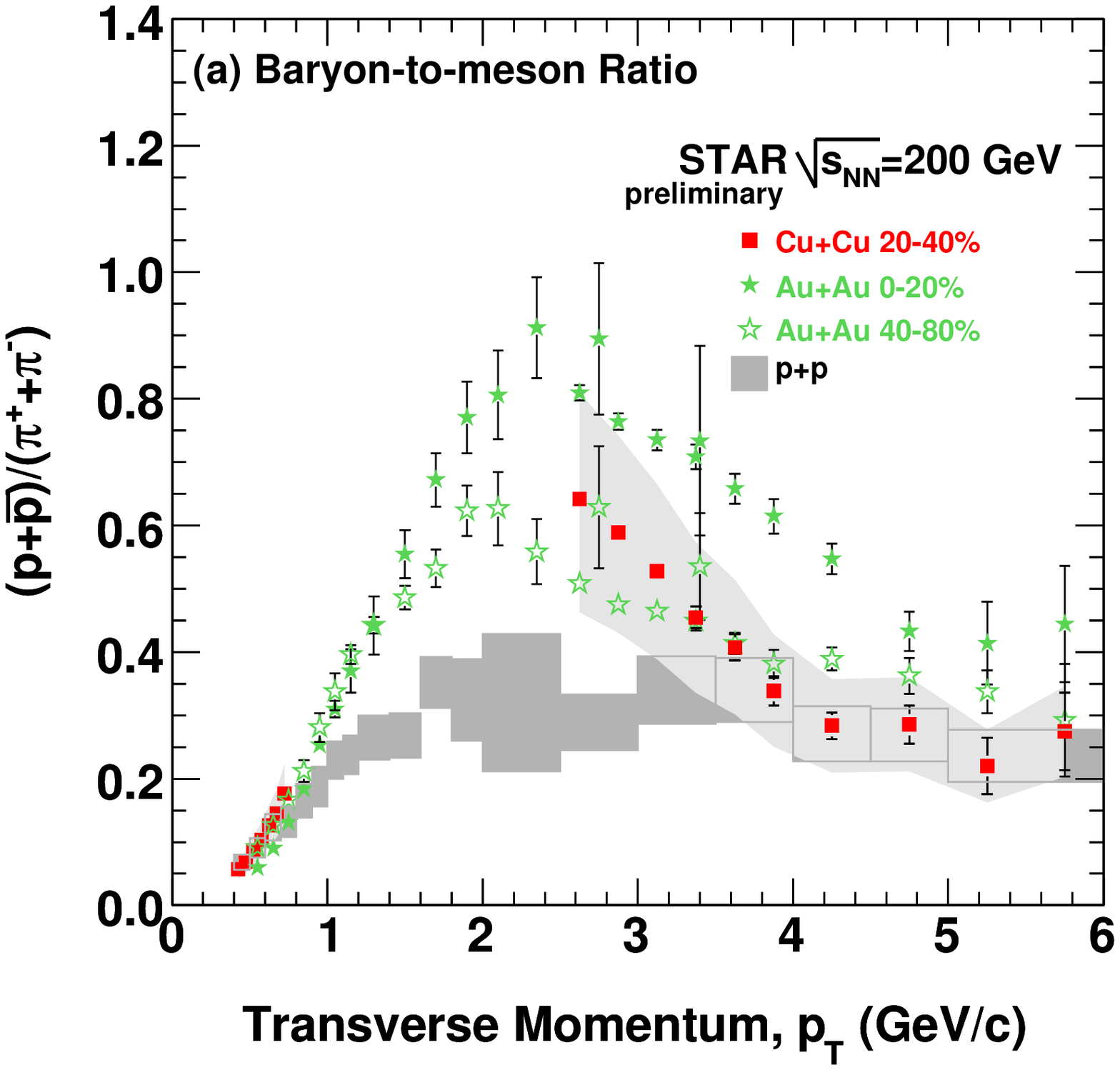}}
\end{minipage}
\hspace{20pt}
\begin{minipage}[t]{.45\textwidth}
\centerline{\epsfxsize=2.4in\epsfbox{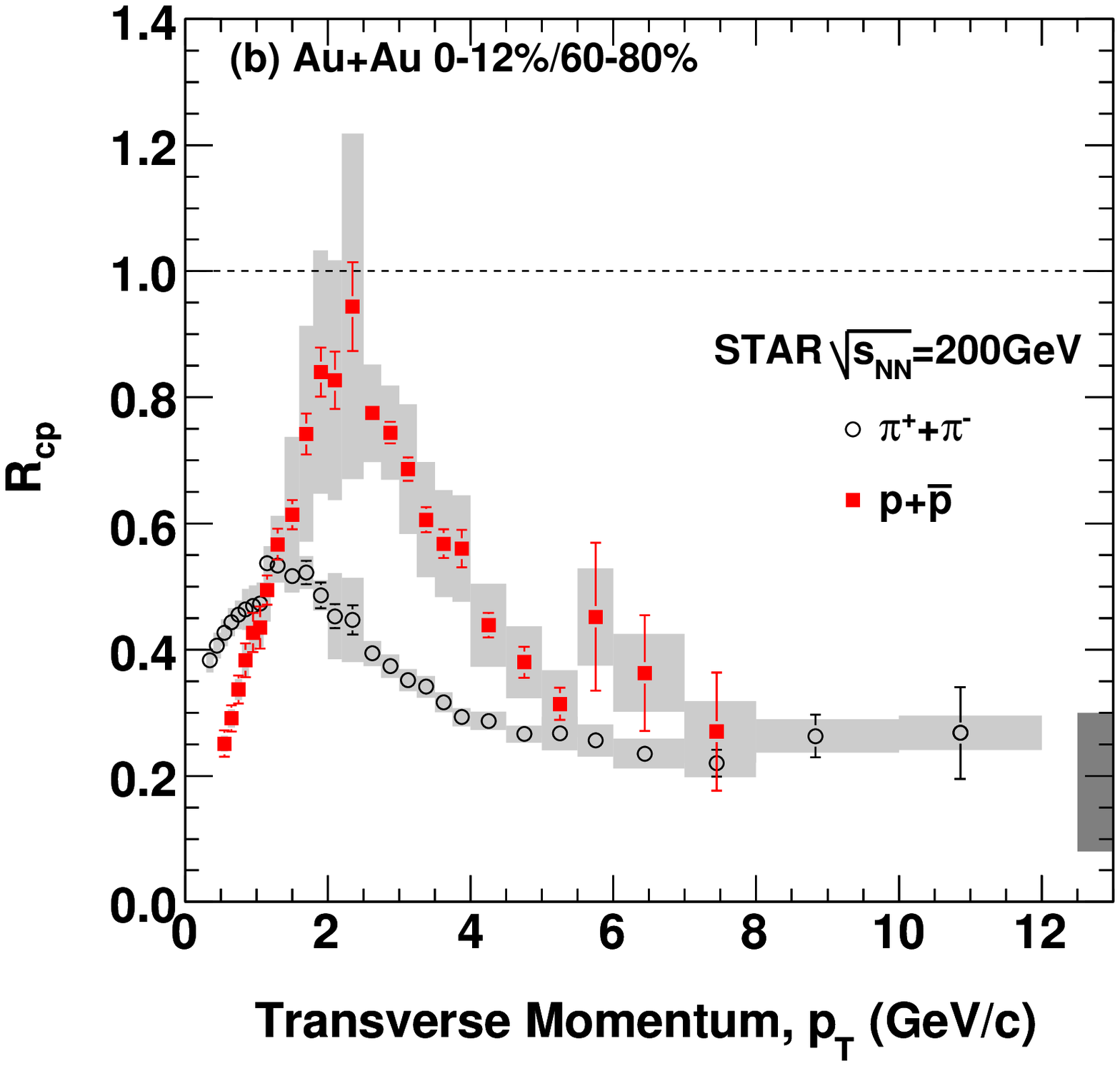}}
\end{minipage}
\caption{Panel (a) shows the transverse momentum dependence of the
baryon-to-meson ($(p+\overline{p})/(\pi^{+}+\pi^{-})$) ratio in Cu+Cu
collisions (squares) at \snn~200~GeV, for the 20-40\% centrality bin, the
light grey band over the data illustrates the systematic uncertainty on the
ratio. For comparison, Au+Au collisions at \snn~200GeV for central (0-20\%,
closed stars) and mid-peripheral (40-80\%, open stars) with statistical
errors only. The mid-peripheral data are
chosen as the mean number of participants matches that of the Cu+Cu data
shown.  Data from p+p collisions are shown as a grey band for reference.
Panel (b) shows the spectral modification (R$_{CP}$) for pions (open circles)
and protons (closed squares) for central (0-12\%) relative to peripheral
(60-80\%) collisions.  The light shaded bands represent the  point to point
systematic uncertainty.  The darker shaded band represents the
normalization systematic uncertainty in the number of binary collisions.
\label{fig:BaryonEnhancement}}
\end{figure}

Another dramatic effect observed in Au+Au data is the relative
enhancement of protons to pions in the intermediate-\pt~region as
compared to p+p and $e^{+}+e^{-}$ collisions~\cite{cite:STAR_Id200AuAu}
and similarly for other baryons and mesons~\cite{cite:STAR_LambdaK0}.
This enhancement is found to be strongly
dependent on the centrality of the collision, as illustrated in
Fig.~\ref{fig:BaryonEnhancement}a.  Peripheral Au+Au data or d+Au
data are found to have baryon to meson ratios similar to that in the
elementary collision systems.  For more central data, the relative number
of baryons produced is found to increase rapidly in the
intermediate-\pt~region, with a peak at \pt$\sim2-3$GeV/$c$.
For high-\pt~particles (\pt$\sim5$GeV/$c$) the enhancement disappears,
with data from all centralities similar to the p+p data.
Similarly to our $\pi$ R$_{AA}$ observations, the baryon to meson ratio
($(p+\bar{p})/(\pi^{+}+\pi^{-})$) is found to be similar in Cu+Cu and
Au+Au collisions for an equivalent number of participants.  Such systematic
agreement between these data points to the system size ($N_{part}$) driving
the distributions.  Although, it has to be pointed out that the number of 
binary collisions for the available centrality bins of Au+Au and Cu+Cu
data are also similar for the same $N_{part}$ values.

\section{Discussion}\label{sec:discussion}

To interpret the presented results one has to look to models for
guidance.  As discussed earlier, it is found that baryons are produced
with a larger contribution from gluon fragmentation than from quark
fragmentation.  It is thus expected that any increase in the baryon
to meson ratio in the intermediate- to high-\pt~range would be resultant
from this source.  To explain the presented data one could consider,
for example, that a gluon jet could be more easily propagated through
the medium than a quark jet, leading to the increase in the number of
protons in the  intermediate-\pt~region.  This, however contradicts
theoretical predictions where an opposite effect was
expected~\cite{cite:Theory_Vitev}.  Alternatively, more gluon jets could
be initially produced, or
{\it induced}, for the more central data. The latter appears to be the
more plausible, as the highest-\pt~data exhibits little or no enhancement
over the p+p data, indicating a similar energy loss for gluons and quarks,
see Fig.~\ref{fig:BaryonEnhancement}.  Alternative approaches to explain
the phenomenon observed in the data, have also been developed.  For example,
the recombination/fragmentation picture of thermal/shower partons has had
success at describing this data~\cite{cite:HwaRecombination}.

Further information on the relative energy loss of quark and gluon jets
can be extracted from the data by comparing the nuclear modification factors
of proton and pion data, Fig.~\ref{fig:BaryonEnhancement}b.  Here we use
the most peripheral Au+Au collisions in the ratio as an approximation
for p+p collisions~\cite{cite:STAR_Id200AuAu}.  As expected from the baryon
to meson ratio, the precise shapes of the R$_{CP}$ is not the same for the
two, most notably in the intermediate-\pt~region.  At high-\pt, however,
the two suppression factors are found to be the same, indicating that the
energy loss of quark and gluon jets may have similar energy loss. 

\begin{figure}[!b]
\begin{minipage}[t]{.45\textwidth}
\centerline{\epsfxsize=2.4in\epsfbox{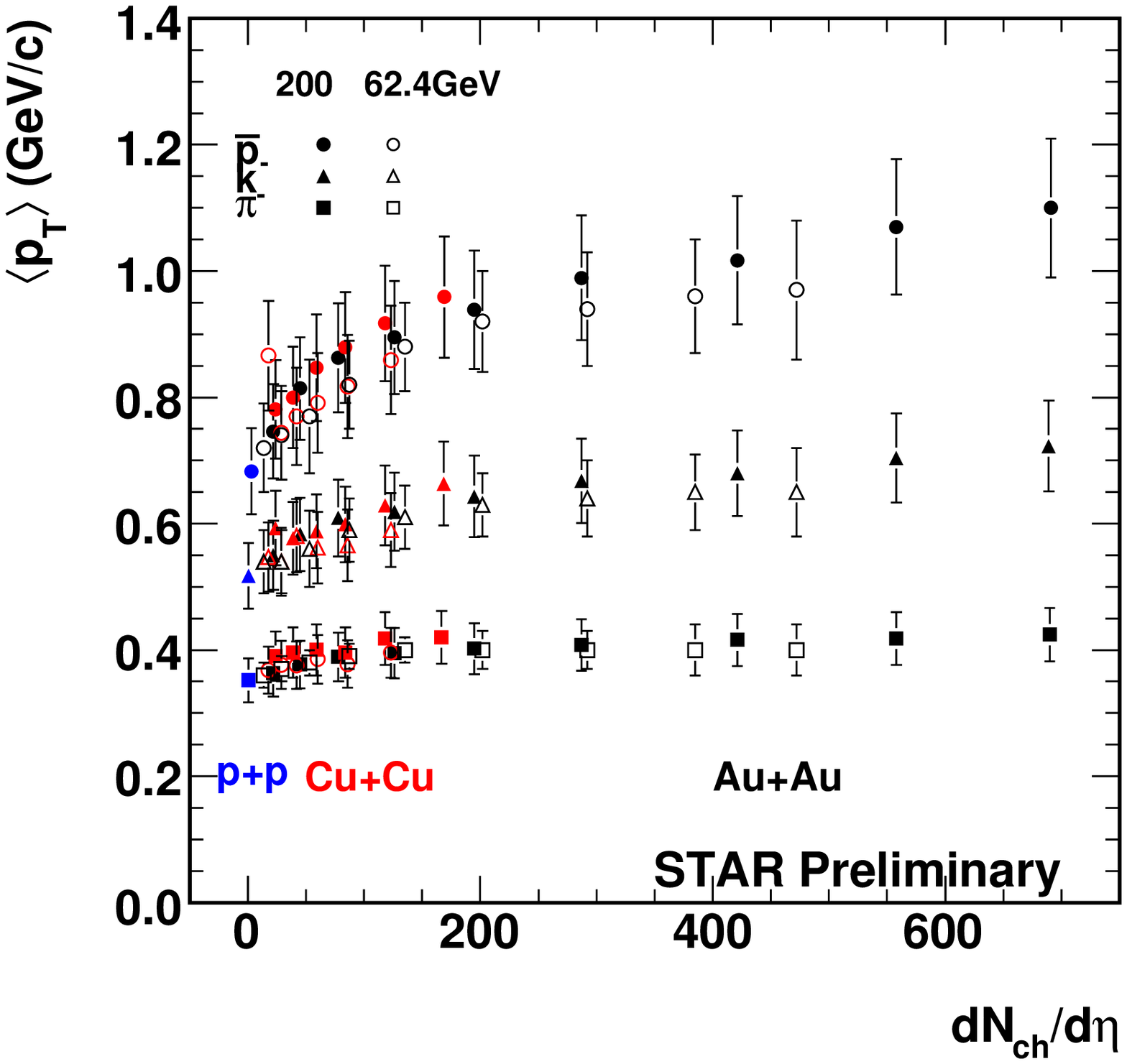}}
\end{minipage}
\hspace{20pt}
\begin{minipage}[t]{.45\textwidth}
\centerline{\epsfxsize=2.4in\epsfbox{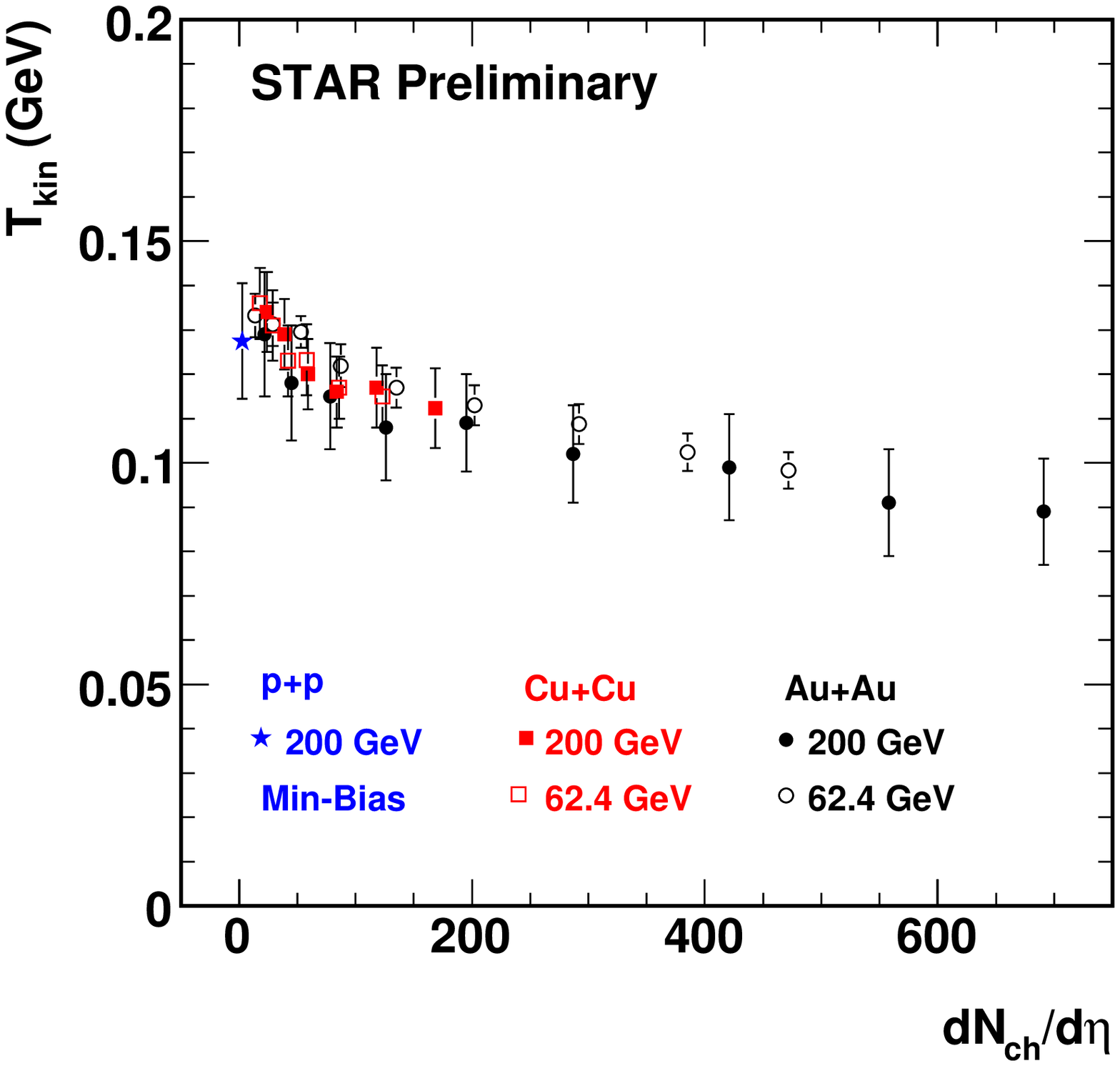}}
\end{minipage}
\caption{Panel (a) shows the centrality ($dN_{ch}/d\eta$) dependence
of the mean-\pt~from p+p to Au+Au collisions, for $\pi^{-}$, K$^{-}$ and
antiprotons.  Panel (b) shows the kinetic freeze-out temperature derived
from blast-wave fits to the low-\pt~spectra.  Data for both panels show
collisions at \snn~200GeV/$c$ for p+p collisions and additionally at
\snn~62.4GeV/$c$ for both Cu+Cu and Au+Au collisions.
\label{fig:Low_pT_vs_Nch}}
\end{figure}

Additional information on the observed enhancement of baryons in Au+Au 
collisions has come from collisions at a reduced center-of-mass energy
of $\sqrt{s_{NN}}=62.4$~GeV.  The relative baryon enhancement is also observed for
this incident energy, although the effect is magnified for the proton
over pion, presumably due to higher baryon transport for this lower
energy.  For the antiproton over pion ratio, the enhancement is lower
due to a smaller number of primordial anti-particles being
produced~\cite{cite:STAR_62Id}.  Although the collision energy is
reduced, a suppression of the high-\pt~hadrons is observed in the nuclear
modification factor.  At this incident energy, the magnitude of the
suppression effect is also increasing as a function of centrality, however
the actual suppression is smaller at $\sqrt{s_{NN}}=62.4$GeV than at
200GeV as compared to the same cross-sectional fraction.  Explanation of
this effect is not trivial as one must consider the fact that the
underlying (p+p) spectrum is softer (lower cross-section for high-\pt).
More detailed discussion on Au+Au collisions at \snn~62.4GeV can be found
in Ref.~\cite{cite:STAR_62Id}.

Systematic effects due to the system size are not limited to the 
high-\pt~regime.  For the bulk particle yields, properties of the chemical
and kinetic freeze-out parameters are found to be strongly coupled to the
geometrical overlap of the system for a given collision energy.  Upon
comparing such parameters additionally for different collision energies,
a consistent picture of bulk particle production comes to the fore.  By
comparing, for example, the mean-\pt~of particle species
(Fig.~\ref{fig:Low_pT_vs_Nch}a), or the derived kinetic freeze-out
temperature, from blast-wave fits (Fig.~\ref{fig:Low_pT_vs_Nch}b), one
observes scaling with the number of charged-hadrons initially produced
in the collisions.  These observations are invariant on the system size, 
collision energy and paint a remarkable picture of particle
production for the whole of the RHIC program.  In a similar fashion to
the high-\pt~results, the Cu+Cu data fit smoothly into the systematics
of the Au+Au data.

\section{Conclusions}

Measurements of identified protons and pions from low- to high-\pt~have
proven to be a valuable tool in understanding the particle production
and energy loss mechanisms in relativistic heavy-ion collisions. The
suppression of pions at high-\pt~lead us to conclude that the partons
undergo a large energy loss due to a hot, dense medium created during the
collisions.  Further studies, through the analysis of protons and pions, 
indicate that the partonic energy loss is similar for both the gluons and
quarks.  The amount of energy loss suffered by the partons is found to
be strongly $N_{part}$ dependent.  For different collision species, the
suppression is found to be invariant for the same number of participants.
Low-\pt~ data show similarly simple scaling behaviors as found at high-\pt.
Mean-\pt~and freeze-out parameters are found to scale along a common curve
for all RHIC energies and collision systems.

\section*{Acknowledgments}

We thank the RHIC Operations Group and RCF at BNL, and the
NERSC Center at LBNL for their support. This work was supported
in part by the Offices of NP and HEP within the U.S. DOE Office 
of Science; the U.S. NSF; the BMBF of Germany; CNRS/IN2P3, RA, RPL, and
EMN of France; EPSRC of the United Kingdom; FAPESP of Brazil;
the Russian Ministry of Science and Technology; the Ministry of
Education and the NNSFC of China; IRP and GA of the Czech Republic,
FOM of the Netherlands, DAE, DST, and CSIR of the Government
of India; Swiss NSF; the Polish State Committee for Scientific 
Research; SRDA of Slovakia, and the Korea Sci. \& Eng. Foundation.

\bibliography{bigsky2007-template}
\bibliographystyle{bigsky2007}
 

\vfill\eject
\end{document}